# Currency Based on Time Standard

Tomas Kala*

*The Total Economic Time Capacity of a Year 525600 minutes as a time standard for a new Monetary Minute currency is postulated in this evaluation study. Consequently, the Monetary Minute ṁ is defined as a 1/525600 part of the Total Economic Time Capacity of a Year. The Value Cṁ of the Monetary Minute ṁ is equal to a 1/525600 part of the GDP, p.c., expressed in a specific state currency C. There is described how the Monetary Minutes ṁ are determined, and how their values Cṁ are calculated based on the GDP and all the population in specific economies. The Monetary Minutes trace different aggregate productivity, i.e. exploitation of the total time capacity of a year for generating of the GDP in economies of different states.*

* Faculty of Informatics and Management, University of Hradec Kralove, Rokitanskeho 62, 530 09 Hradec Kralove, the Czech Republic (e-mail: tomas.kala@uhk.cz or t.kala@centrum.cz).

## I. Introduction

The economics has been lacking an appropriate tool for determining and measuring the value of economic variables for ages. There are numerous theories of value, like

- the theory of Intrinsic and extrinsic Value (Zimmerman and Bradley, 2019)
- the Labor theory of value (Junankar, 1982), (Peach, 1993)

- the Exchange theory of value (Rubin, 2008)
- the Monetary theory of value (Milios, 2012)
- the Power theory of value (Nitzan & Bichler, 2009)
- the Subjective theory of value (Kagan, 2019.)
- the Marginal Utility (Bloomenthal, 2019.)
- The Wealth of Nations (Smith, 2018)
- the Theory of Value (Economics), (Collective of Authors, 2019).

Nevertheless, it was money, which played a role of the Value in applied economy, entrepreneurship, business, everyday human activities for ages - despite the fact that they show pros and cons (Abel, Bernanke, and Croushore 2015; Samuelson and Nordhaus 2010; Mankiw 2010; Maurer 2006; M Acemoglu and Robinson 2012; Priest 2001; Smith 2018; Meltzer and Friedman n.d.; Junankar 1982). Money has very significant functions for economic activities, namely as a medium of exchange, unit of account, store of value (Abel et al., 2015; Samuelson & Nordhaus, 2010), and unfortunately as a medium for speculation and crime (Biggs 2016; Porter and Rousse 2016).

Money has got various forms, as a commodity physical money (such as shells, corals, stones, pearls, gold, silver), a representative money (i.e. paper money, banknotes, gold standard notes, coins, counterfeit money, fiat money, bank money) alternative or complementary money currencies based on time spent in producing services or goods (e.g. the Time Dollar, Time Credits, Service Credits). (See: Abel, Bernanke, and Croushore 2015; Samuelson and Nordhaus 2010; Mankiw 2010; Maurer 2006; Acemoglu and Robinson 2012; Priest 200; Meltzer and Friedman 2019; Cahn and Rowe 1992). Special category of current money creates digital money like Bitcoin, and many others Math-Based Money (MBC), which are not backed fiat-based currencies (Biggs, 2016; Porter & Rousse, 2016).

Values of all of the type of money have been based on negotiation, bargaining, settlement, agreement, market situation, economic, and political power or position of the participants - above all on the decisions of the monetary policymakers, and on negotiation ability and power of individual merchants, firms, companies, countries, banks, public, etc. ((Abel, Bernanke, and Croushore 2015; Samuelson and Nordhaus 2010; Mankiw 2010). The typical monetary policymakers are sovereign rulers, governments, central bankers, namely Federal Reserve System, which consists of 12 Federal Reserve banks, together with the Board of Governors in Washington, D.C., in the U.S.A.).

A special role had played gold, silver, and some other commodities, which used to serve as monetary standards in the value set in economies up to 1973 (Abel, Bernanke, and Croushore 2015; Samuelson and Nordhaus 2010; Mankiw 2010). However, the availability, amount, and rareness of these commodities – and consequently a variety of their value - have fluctuated over time.

The problematics of the commodity standard setting were discussed in Milton Friedman´s book (Friedman & Friedman, 1982), as well as in Hayek´s work (Hayek, 1990).

The problem of the Money Value setting has been even magnified by the fact, that modern monetary systems are prevailingly debt-based (enabling fractional reserve banking) instead of Value-based (Abel et al., 2015; Mankiw, 2010; Samuelson & Nordhaus, 2010). Furthermore, the amount of money in current economies - and consequently its Value – are set subjectively by governments and central banks and/or by other monetary authorities and theirs declared monetary policy. These two aspects contribute/cause devastating financial crisis from time to time (Angelides Commission 2011; Bernanke 2009; Classens and Kose 2013; Stiglitz and et.al. 2009; Stiglitz 2010).

To sum-up: modern money is not tied - at the time being - to any “firm” or material (like gold), or any other background which would be precious, predictably evolving independently on any subjective power (Abel, Bernanke, and Croushore 2015; Samuelson and Nordhaus 2010; Mankiw 2010).

In my previous works (Kala 2014, 2016, 2017, 2018, 2019), I have postulated Time as a Commodity for establishment, measurement, and evaluating the value of economic entities and processes. I have defined a new currency called the "Time-Based Money" (in abbreviation the TBM), (defined as a value of the GDP per capita divided by the number of minutes per year (i.e. by 525600) (Kala 2016, 2017, 2018, 2019).

I have tested the Time-Based Money values in the U.S. and the Czech economies (Kala, 2016), in the Slovak, Polish, and Hungary economies, for 2011-2015 years (Kala, 2017). I have revealed - among others features - that the TBM values expressed in relevant national currencies exhibited a systematic increase in their values. However, the TBM values based on U.S.$ and Euro currencies showed erratic development in the period studied without any apparent objective reasons (Kala, 2018).

In my last evaluation study, I have focused on the comparison of the living costs and/or incomes evaluated in so-called Monetary Minute currencies (further in abbreviation MonMin or ṁ), which were related to the currently utilized currencies (like in $, €, £, and others) via the above mentioned TMB (Kala, 2019).

In this article, I am going to explain reasons why time should be accepted as a monetary standard, refine the terminology used in my design of the Currency Based on Time Standard, and present results on testing the usability of the MonMin currency in economies, monetary policies, financial operations, etc.

## II. Why Time should be accepted as a Monetary Standard

Time can be supposed as one of the specific entities, which our world consists of (Kala, 2014b); its value is extraordinary per se.

Time - namely the human life-time - is one the most precious value in the human world.

Each of us has at disposal just 24 hours or 1440 minutes, or 86400 seconds total in a day; or approximately 525600 minutes a year.[1]

Time is not a subject of inflation, for an hour in 3600 seconds today, and will be tomorrow and in the future. Especially this feature of time can serve as a firm "standard", or "commodity", for a valuable time-steady monetary system.

An amount of the time of a country or a state is in a direct correlation with an increase or decrease of its population, which could be used as a useful measure or as a natural regulator for the money supply rules of a currency generation in national or international economies.

The quantity "Time" is commensurate with economics, its theory and practice, for it allows to define numerous economic entities like money as a store of value, money velocity, productivity of human labor, depreciation of assets with time, interest on money, inflation of money, etc.

The amount of the time total is at disposal to countries or states (or broadly speaking to mankind) for managing all the emerging challenges, possibilities, opportunities, threads, and other circumstances coupled with the increasing or decreasing number of population on Earth.

A currency based on the time standard might reflect, differ, and measure a quality of managerial processes, effectivity, efficiency, productivity, as well as an influence of geographical, demographical, nature, political, social, and others factors in different places and historical on real economic processes in "homogenous" units, i.e. in time units.

[1] There is 525948.766 minutes total in a year (Google, n.d.)

## III. Methodology

This evaluation study is based on the method of qualitative analysis of secondary data and information available on the function of money.

Namely, the introductory survey on the state of the art given in Chapter I. has been done using database mining in the Web of Science Core Collection, SCOPUS, and ProQuest databases and other resources. The references related to the information resources used were sorted using the Mendeley Reference Manager produced by Elsevier.

In this work, I have postulated a time standard for a new Monetary Minute (MonMin, or ṁ)[2] currency as the total Economic Time Capacity of a Year 525600 minutes[3].

Consequently, I have defined the Monetary Minute ṁ as a 1/525600 part of the total Economic Time Capacity of a Year.

The Monetary Minute ṁ has a Value of $C\dot{m}$, which is equal to a 1/525600 part of the GDP, p.c., expressed in a specific state currency C. It can be calculated as follows:

(1) $C\dot{m}$ = GDP / NP / TETCY, where

$C\dot{m}$ denotes the Value of the Monetary Minute Currency unit (in $C/\dot{m}$), i.e. the Value of the 1/525600 part of the GDP, p.c., expressed in a specific state currency C.

NP is for Number of Population,

GDP, p.c. is the Gross Domestic Product per Capita,

TETCY is a Total Economic Time Capacity of a Year expressed in

[2] Minutes is a unit used as a measure of time usually, however, the Monetary Minute ṁ is a unit used – in my conception –as a measure of an economic Value.

[3] Note: A Labor Time – in an understanding of the Political Economy – which is spent while creating the GDP is included as a relatively small part in the total Economic Time Capacity of a Year.

Monetary Minutes ṁ (i.e. approximately 365 days x 24 hours x 60 minutes = 525600 ṁ).

That means, that one Monetary Minute ṁ (i.e. the 1/525600 part of the total Economic Time Capacity of a Year) has a Value of C*ṁ*, which is equal to a 1/525600 part of the GDP, p.c., expressed in a specific state currency C.

The C*ṁ* values of selected currencies C (see Table 2) were calculated as well as:

(2) $C\dot{m} = \$\dot{m} \times \text{CurrencyRate}$, where

C*ṁ* denotes the Value of a Monetary Minute ṁ in a specific Currency C

$ṁ is a Cṁ Value determined from the GDP and a Population of The U.S.A, and the Total Economic Time Capacity of a Year Time, i.e. 525600 ṁ according to the Equation (1)

Using the Monetary Minute Values, one can calculate prices of products and services or values of incomes, salaries, costs, and other quantities in Monetary Minutes as:

(3) $P_M = P_C / C\dot{m}$, where

$P_M$ denotes a Price of a Product or a Service or other quantities in Monetary Minutes, i.e. *ṁ*,

$P_C$ denotes a Price of a Product or a Service or other quantities in a specific Currency C, in which the GDP or the GDP, p.c. – and consequently – the Monetary Minutes *ṁ* are expressed,

C*ṁ* denotes the Value of a MonMin in a specific Currency C[4].

[4] Note: The nominal values of the Cṁ currencies – and consequently the nominal values of the prices of products or services or other quantities expressed in Monetary Minutes ṁ - will differ reflecting the Currencies in which the GDP or the GDP p.s. are expressed.

The values of the GDP and the Number of Population of the U.S. were taken from the (U.S.BureauofEconomic, 2019; U.S.CensusBureau 2019; OfficeforNationalStatistics 2019; StatistischeÄmterdesBundesundLänder 2019a; StatistischeÄmterdesBundesundLänder 2019; NationalBureauofStatisticsofChina 2018; Japanmacroadvisors 2019b; Japanmacroadvisors 2019a; CzechStatisticalOffice 2019; CzechStatisticalOffice 2019).

The historical data on M1 and GDP p.c. related to the U.S. economy were found on the U.S. Bureau of Labor Statistics web sites (FinanciaForecastCenter, 2019).

I used values of Living Costs and Incomes in several countries taken on July 1$^{st}$, 2019 from NUMBEO (2019).

I used currency ratios of several Currencies from the Trading Economy, Kurzy, CZ, and FOREX, (Forex, 2019; Kurzycz, 2019.; TradingEconomics, 2019). Commodity rates were used from Kurzy CZ (Kurzycz, 2019).

List of the MonMin Currency was inspired by the XE Currency Encyclopedia style of the USD currency list (Corporation, 2019).

## IV. Results

The Value of the Monetary Minute for the U.S. $\$\dot{m}$ (at Jan 1$^{st}$, 2019) was calculated according to Equation 1 as follows:

$$\$\dot{m} = 20891.4 / 328467812 / 525600$$
$$= 0.1210095 \text{ \$ per MonMin, where}$$

| | |
|---|---|
| 20,891.4 | is the GDP of the U.S.A. at the 4$^{th}$ Q 2018 in billions current $ (TheU.S.BureauofEconomicAnalysis, 2019), |
| 328.467,812 | is the number the Resident Population Plus Armed Forces Overseas the U.S.A., Jan 1$^{st}$, 2018 (U.S.CensusBureau 2019), |
| 525600 | is the Economic Time Capacity of a Year expressed in MonMin |

Analogously were calculated Cṁ values for the United Kingdom, Germany, China, Japan, and the Czech Republic currencies (see Table 1).

[ Insert Table 1 Here]

TABLE 1—THE VALUES OF THE MONETARY MINUTES *Cṁ* CALCULATED ON THE BASE OF THE GDP VALUES AND THE NUMBERS OF POPULATION FOUND IN NATIONAL DATABASES OF THE U.S.A., GERMANY, UNITED KINGDOM, CHINA, AND THE CZECH REPUBLIC

| State | GDP | | Population | GDP, p.c. | | *Cṁ* | |
|---|---|---|---|---|---|---|---|
| The U.S.A. | 20891400000000 | $ | 328467812 | 63603 | $ | 0.1210095 | US$/MonMin |
| The U.K. | 511482000000 | £ | 66435600 | 7699 | £ | 0.0146479 | £/MonMin |
| Germany | 3386000000000 | € | 82887000 | 40851 | € | 0.0777222 | €/MonMin |
| China | 253598600000000 | ¥ | 13900800000 | 18243 | ¥ | 0.0347098 | ¥/MonMin |
| Japan | 549700000000000 | Y | 126200000 | 4355784 | Y | 8.2872612 | Y/MonMin |
| The Czech Republic | 5328738000000 | Kč | 10649800 | 500360 | Kč | 0.9519794 | Kč/MonMin |

*Source:* Author calculations using data from: (U.S.BureauofEconomic, 2019; U.S.CensusBureau 2019; OfficeforNationalStatistics 2019; StatistischeÄmterdesBundesundLänder 2019a; StatistischeÄmterdesBundesundLänder 2019; NationalBureauofStatisticsofChina 2018; Japanmacroadvisors 2019b; Japanmacroadvisors 2019a; CzechStatisticalOffice 2019; CzechStatisticalOffice 2019).

It is obvious from Table 1, that the Values of the Monetary Minutes related to specific states were very different. These phenomena were obviously caused by different values of the GDP, numbers of Population, consequently values of the GDP, p.c., different state currencies values, and as well by different “aggregate productivity”, i.e. different exploitation of the Total Time Capacity of a Year for generating of the GDP in Economies of individual states.

In Table 2, there are given the Cṁ rates related to the several currencies, based on the exchange rates of the Trading Economics (at July 1st, 2019) (TradingEconomics, 2019). The displayed values of Cṁ were calculated according to Equation 2.

TABLE 2—THE MONMIN RATES RELATED TO THE SEVERAL CURRENCIES BASED ON THE EXCHANGE RATES OF THE TRADING ECONOMICS (AT JULY 1ST, 2019)

| | $ | € | £ | Y | ¥ |
|---|---|---|---|---|---|
| $ | 1 | 1.1325 | 1.26537 | 0.00923276 | 0.14607 |
| C*ṁ*, in $/MonMin | 0.11918 | 0.134971 | 0.170789 | 0.00157685 | 0.00023033 |
| Inverse C*ṁ*, in MonMin/$ | 8.39 | 7.41 | 5.86 | 634.18 | 4341.59 |

*Source:* Author calculations using data from: (TradingEconomics 2019).

It is obvious from Table 2, that one $ṁ had a smaller nominal value (0.11918) than one €ṁ (0.134971), and also than one £ṁ (0.170789). On the contrary, 1 Yṁ, as well as 1 ¥ṁ, had their nominal values (0.00923276, and of 0.146070, respectively) much smaller, hence much shorter MonMin values (just 0.00157685 MonMin, and 0.000230330 MonMin) than 1 $ṁ has. The content of Table 2 can also be understood that 1 $ represented 8.39 MonMin of Time, 1 € corresponded to 7.41 MonMin, 1 £ corresponded to 5.86, 1 Y was adequate to 634.18 MonMin, and 1 ¥ corresponded to 4341.59 of MonMin, respectively, in the financial market of the United States.

The money market values of related Cṁ and the calculated values based on the GDP of the economies included in both Tables 1, and 2 were significantly different.

In Table 3, there is given a list of several commodity prices in the Czech market, at randomly selected day July 18th, 2019 (Kurzycz, 2019). They were expressed in U.S. dollars, Czech currency (CZK), EUR, Great Britain Pounds, and also in relevant Monetary Minute values, respectively. It can be seen, that prices of individual commodities expressed in related currencies were very different and confusing. Meanwhile, the ones expressed in the related Monetary Minute currencies enabled to compare and testify their values in the Czech market concerning the currency market in the Czech Republic.

TABLE 3—LIST OF SEVERAL COMMODITY PRICES IN THE CZECH COMMODITY MARKET EXPRESSED IN $, CZK, €, £ CURRENCIES, AND IN RELATED MONMIN VALUES (AT JULY 18TH, 2019) (Kurzycz, n.d., 2019)

| Commodity | Unit | Price in Traditional Currencies | | | | Price in MonMin Currencies | | | |
|---|---|---|---|---|---|---|---|---|---|
| | | $ | CZK | € | £ | US MonMin | CZE MonMin | GER MonMin | UK MonMin |
| Electricity* | 1 MWh | 67.82 | 1416.69 | 55.45 | 49.87 | 560 | 1488 | 713 | 3404 |
| Crude Oil Brent | 1 Barel | 62.43 | 1421.22 | 55.63 | 50.02 | 516 | 1493 | 716 | 3415 |
| Natural Gas | MMBtu | 2.29 | 52.13 | 2.04 | 1.83 | 19 | 55 | 26 | 125 |
| Gold | 1 oz | 1447.00 | 32940.96 | 1289.32 | 1159.46 | 11958 | 34603 | 16589 | 79155 |
| Wheat | 100 Bushl | 493.50 | 11234.53 | 439.72 | 395.43 | 4078 | 11801 | 5658 | 26996 |
| Cotton | 100 Pound | 61.71 | 1404.83 | 54.99 | 49.45 | 510 | 1476 | 707 | 3376 |
| Currency Ratio | CZK** | 22.765 | 1 | 25.549 | 28.409 | | | | |
| Cṁ, in C/MonMin*** | | 0.121001 | 0.951979 | 0.077722 | 0.014648 | | | | |

*Source: Author´s calculations using data from the Kurzy.cz* (Kurzycz, 2019)

* *on July 15th, 2019*

** *on July 18th, 2019*

****See Table 1*

In Table 3, a noticeable apparent overvaluation of the U.S. dollar against the values of other currencies in the Czech Republic at that time can be derived from the MonMin values, as the nominal values of one ounce of gold in the Czech commodity market were $P_{1OZGold}$ = 11958 $ṁ, 34603, $P_{1OZGold}$ = 34603 Kčṁ, $P_{1OZGold}$ = 16589 €ṁ, and $P_{1OZGold}$ = 79155 £ṁ, respectively. The equalization of the nominal Values of the prices of one ounce of gold in the Czech commodity market to a nominal Value of $P_{1OZGold}$ = 11958 $ṁ would be possible under a hypothetical presumption that the currency conversion ratios of the Czech Crown should be 7.867 Kč/$, 18.417 Kč/€, and 4.292 Kč/£, respectively (at July 18th, 2019).

In Table 4, there is given a list of several food staff costs and also salaries expressed in the Monetary Minute Currencies Cṁ values in several states.

TABLE 4—LIST OF SEVERAL FOOD STAFFS PRICES AND AVERAGE MONTHLY DISPOSABLE SALARIES EXPRESSED IN THE CṀ VALUES IN COUNTRIES UNDER STUDY ON JULY 1ST, 2019.

| Item | The U.S.A. | Germany | The U.K. | Japan | China | Czech |
|---|---|---|---|---|---|---|
| | $ṁ | €ṁ | £ṁ | Yṁ | ¥ṁ | Kčṁ |
| Restaurants | | | | | | |
| Meal, Inexpensive Restaurant | 123 | 129 | 825 | 109 | 720 | 137 |
| Meal for 2 People, Mid-range Rest., 3-course | 413 | 579 | 3162 | 483 | 4322 | 630 |
| McMeal at McDonalds (or Equivalent Meal) | 58 | 96 | 375 | 82 | 922 | 137 |
| Domestic Beer (0.5 liter draught) | 33 | 45 | 241 | 48 | 202 | 37 |
| Imported Beer (0.33 liter bottle) | 50 | 41 | 247 | 60 | 435 | 42 |
| Cappuccino (regular) | 34 | 34 | 181 | 48 | 779 | 46 |
| Coke/Pepsi (0.33 liter bottle) | 15 | 29 | 85 | 16 | 96 | 30 |
| Water (0.33 liter bottle) | 12 | 24 | 63 | 13 | 61 | 22 |
| Markets | | | | | | |
| Milk (regular), (1 liter) | 7 | 9 | 63 | 22 | 375 | 19 |
| Loaf of Fresh White Bread (500g) | 21 | 17 | 67 | 23 | 315 | 24 |
| Rice (white), (1kg) | 32 | 27 | 61 | 60 | 202 | 38 |
| Eggs (regular) (12) | 19 | 24 | 128 | 27 | 364 | 44 |
| Local Cheese (1kg) | 86 | 115 | 371 | 201 | 2569 | 199 |
| Chicken Breasts (Boneless, Skinless), (1kg) | 70 | 94 | 395 | 93 | 724 | 148 |
| Beef Round (1kg) (or Equivalent) | 95 | 143 | 516 | 285 | 1970 | 236 |
| Apples (1kg) | 37 | 28 | 124 | 90 | 361 | 32 |
| Banana (1kg) | 12 | 20 | 74 | 37 | 219 | 32 |
| Oranges (1kg) | 33 | 27 | 115 | 68 | 341 | 36 |
| Tomato (1kg) | 33 | 32 | 119 | 73 | 236 | 45 |
| Potato (1kg) | 21 | 12 | 79 | 49 | 169 | 18 |
| Onion (1kg) | 21 | 14 | 65 | 40 | 189 | 17 |
| Lettuce (1 head) | 13 | 12 | 51 | 24 | 129 | 23 |
| Water (1.5 liter bottle) | 15 | 4 | 67 | 15 | 120 | 13 |
| Bottle of Wine (Mid-Range) | 99 | 64 | 481 | 157 | 2161 | 125 |
| Domestic Beer (0.5 liter bottle) | 17 | 10 | 109 | 31 | 147 | 16 |
| Imported Beer (0.33 liter bottle) | 23 | 16 | 126 | 40 | 378 | 29 |
| Cigarettes 20 Pack (Marlboro) | 59 | 82 | 687 | 57 | 576 | 105 |
| Salaries And Financing | | | | | | |
| Average Monthly Net Salary (After Tax) | 25867 | 28464 | 122552 | 34409 | 173308 | 25501 |
| Cṁ Values used (see Table 1) | 0.12101 | 0.07772 | 0.014548 | 8.2873 | 0.0347 | 0.95198 |
| Currency Coversion Rates used | 1.00 | 0.88 | 0.79 | 108.33 | 6.85 | 22.44 |

*Source: Author´s calculations using data from NUMBEO* (NUMBEO, 2019f, 2019c, 2019e, 2019d, 2019a, 2019b)

The nominal values of individual items of the living costs expressed in Cṁ showed large and confused disparities both, in expenditures, and salaries.

The MonMin prices of the McMeal at McDonald could be recalculated into the nominal value of the Unites State MonMin, i.e. $P_{McMeal}$ = 58 $ṁ, under a hypothetical presumption that the currency conversion ratios would be 0.53 €/$, 0.12 £/$, 76.62 Y/$, 0.43 ¥/$, and 9.50 Kč/$.

In Table 4b., there is given a List of Several Food Staffs Prices and Average Monthly Disposable Salaries Expressed in percent of the Average Monthly Net Salaries in the countries under study in 2019.

TABLE 4B—LIST OF SEVERAL FOOD STAFFS PRICES AND AVERAGE MONTHLY DISPOSABLE SALARIES EXPRESSED IN PERCENT OF THE AVERAGE MONTHLY NET SALARIES IN THE COUNTRIES UNDER STUDY ON JULY 1ST, 2019.

| Item | The U.S.A. percent of Net Salary | Germany percent of Net Salary | The U.K. percent of Net Salary | Japan percent of Net Salary | China percent of Net Salary | Czech Rep. percent of Net Salary |
|---|---|---|---|---|---|---|
| Restaurants | | | | | | |
| Meal, Inexpensive Restaurant | 0.48 | 0.45 | 0.67 | 0.32 | 0.42 | 0.54 |
| Meal for 2 People, Mid-range Rest., 3-course | 1.60 | 2.03 | 2.58 | 1.40 | 2.49 | 2.47 |
| McMeal at McDonalds (or Equivalent Meal) | 0.22 | 0.34 | 0.31 | 0.24 | 0.53 | 0.54 |
| Domestic Beer (0.5 liter draught) | 0.13 | 0.16 | 0.20 | 0.14 | 0.12 | 0.14 |
| Imported Beer (0.33 liter bottle) | 0.19 | 0.14 | 0.20 | 0.18 | 0.25 | 0.16 |
| Cappuccino (regular) | 0.13 | 0.12 | 0.15 | 0.14 | 0.45 | 0.18 |
| Coke/Pepsi (0.33 liter bottle) | 0.06 | 0.10 | 0.07 | 0.05 | 0.06 | 0.12 |
| Water (0.33 liter bottle) | 0.05 | 0.09 | 0.05 | 0.04 | 0.04 | 0.09 |
| Markets | | | | | | |
| Milk (regular), (1 liter) | 0.03 | 0.03 | 0.05 | 0.06 | 0.22 | 0.07 |
| Loaf of Fresh White Bread (500g) | 0.08 | 0.06 | 0.05 | 0.07 | 0.18 | 0.09 |
| Rice (white), (1kg) | 0.12 | 0.09 | 0.05 | 0.17 | 0.12 | 0.15 |
| Eggs (regular) (12) | 0.07 | 0.08 | 0.10 | 0.08 | 0.21 | 0.17 |
| Local Cheese (1kg) | 0.33 | 0.40 | 0.30 | 0.58 | 1.48 | 0.78 |
| Chicken Breasts (Boneless, Skinless), (1kg) | 0.27 | 0.33 | 0.32 | 0.27 | 0.42 | 0.58 |
| Beef Round (1kg) (or Equivalent) | 0.37 | 0.50 | 0.42 | 0.83 | 1.14 | 0.92 |
| Apples (1kg) | 0.14 | 0.10 | 0.10 | 0.26 | 0.21 | 0.13 |
| Banana (1kg) | 0.05 | 0.07 | 0.06 | 0.11 | 0.13 | 0.12 |
| Oranges (1kg) | 0.13 | 0.10 | 0.09 | 0.20 | 0.20 | 0.14 |
| Tomato (1kg) | 0.13 | 0.11 | 0.10 | 0.21 | 0.14 | 0.18 |
| Potato (1kg) | 0.08 | 0.04 | 0.06 | 0.14 | 0.10 | 0.07 |
| Onion (1kg) | 0.08 | 0.05 | 0.05 | 0.12 | 0.11 | 0.07 |
| Lettuce (1 head) | 0.05 | 0.04 | 0.04 | 0.07 | 0.07 | 0.09 |
| Water (1.5 liter bottle) | 0.06 | 0.02 | 0.05 | 0.04 | 0.07 | 0.05 |
| Bottle of Wine (Mid-Range) | 0.38 | 0.23 | 0.39 | 0.46 | 1.25 | 0.49 |
| Domestic Beer (0.5 liter bottle) | 0.07 | 0.03 | 0.09 | 0.09 | 0.08 | 0.06 |
| Imported Beer (0.33 liter bottle) | 0.09 | 0.05 | 0.10 | 0.12 | 0.22 | 0.11 |
| Cigarettes 20 Pack (Marlboro) | 0.23 | 0.29 | 0.56 | 0.17 | 0.33 | 0.41 |
| Salaries And Financing | | | | | | |
| Average Monthly Net Salary (After Tax) | 100.00 | 100.00 | 100.00 | 100.00 | 100.00 | 100.00 |

*Source: Author´s calculations using data from NUMBEO* (NUMBEO, 2019f, 2019c, 2019e, 2019d, 2019a, 2019b)

It is my opinion, that the numbers in Table 4 and Table 4b can be understood as a gauge (a measure) for measurement the relations between costs of products and a consumption of a human lifetime, as the values of costs are expressed in time units (MonMin) or in percentage of the salaries (which are also expressed in time units MonMin or in percent), which people receive as a reward (a value) for their time they (hypothetically or statistically) spent, while generating their ideal share in the General Domestic Product. It can be noticed from the figures in the Table 4b, that the relative prices for meals and drinks in the countries under study were the same in order of several tenths to units of percent in 2019.

In Table 5, there are given values of Money Stock M1 in billions of $, M1 in billions of $ṁ, GDP in billions of $, and Events which affected the GDP in the U.S.A. in 1960-2016. The values were used in Figure 1, in which the time development of the individual items are visualized.

TABLE 5—VALUES OF MONEY STOCK M1 IN BILLIONS OF $, M1 IN BILLIONS OF $Ṁ, GDP IN BILLIONS OF $, AND EVENTS, WHICH AFFECTED THE GDP IN THE U.S.A. FROM 1960 TO 2016.

| Year | M1 in Billions $ | M1 in Billions $ṁ | GDP in Billions $ | Events Affecting GDP |
|---|---|---|---|---|
| 1960 | 140 | 24529 | 542 | Recession. |
| 1961 | 141 | 24240 | 562 | JFK's deficit spending ended recession. |
| 1962 | 145 | 23570 | 604 | |
| 1963 | 148 | 23120 | 638 | |
| 1964 | 154 | 22630 | 685 | LBJ's Medicare, Medicaid. |
| 1965 | 161 | 22118 | 742 | |
| 1966 | 169 | 21488 | 813 | Vietnam War. |
| 1967 | 172 | 20876 | 860 | |
| 1968 | 184 | 20661 | 941 | Moon landing. |
| 1969 | 199 | 20793 | 1018 | Nixon took office. |
| 1970 | 206 | 20711 | 1073 | Recession. |
| 1971 | 216 | 20190 | 1165 | Wage-price controls. |
| 1972 | 230 | 19847 | 1279 | Stagflation. |
| 1973 | 252 | 19657 | 1425 | End of gold standard. |
| 1974 | 264 | 19192 | 1545 | Watergate. |
| 1975 | 274 | 18452 | 1685 | Recession ended. |
| 1976 | 288 | 17618 | 1876 | Fed lowered rate. |
| 1977 | 308 | 17141 | 2082 | |
| 1978 | 334 | 16633 | 2352 | Fed raised rate to 20% to stop inflation. |
| 1979 | 359 | 16147 | 2627 | |
| 1980 | 386 | 16127 | 2857 | Recession. |
| 1981 | 411 | 15468 | 3207 | Reagan tax cut. |
| 1982 | 443 | 16120 | 3344 | Recession ended. |
| 1983 | 477 | 16136 | 3634 | Tax hike and defense spending. |
| 1984 | 525 | 16118 | 4038 | |
| 1985 | 557 | 16053 | 4339 | |
| 1986 | 621 | 17124 | 4580 | Tax cut. |
| 1987 | 730 | 19153 | 4855 | Black Monday. |
| 1988 | 756 | 18560 | 5236 | Fed raised rates. |
| 1989 | 786 | 18066 | 5642 | S&L Crisis. |
| 1990 | 795 | 17490 | 5963 | Recession. |
| 1991 | 827 | 17803 | 6158 | Recession. |
| 1992 | 910 | 18717 | 6520 | NAFTA drafted |
| 1993 | 1030 | 20354 | 6859 | Balanced Budget Act. |
| 1994 | 1132 | 21248 | 7287 | |
| 1995 | 1151 | 20817 | 7640 | Fed raised rate. |
| 1996 | 1124 | 19401 | 8073 | Welfare reform. |
| 1997 | 1081 | 17740 | 8578 | |
| 1998 | 1074 | 16833 | 9063 | LTCM crisis. |
| 1999 | 1098 | 16342 | 9631 | Repeal of Glass-Steagall. |
| 2000 | 1122 | 15826 | 10252 | Tech bubble burst. |
| 2001 | 1097 | 15116 | 10582 | 9/11 attacks. |
| 2002 | 1190 | 16016 | 10936 | War on Terror. |
| 2003 | 1227 | 15892 | 11458 | Iraq War. JGTRRA. |
| 2004 | 1306 | 16003 | 12214 | |
| 2005 | 1367 | 15835 | 13037 | Katrina. Bankruptcy Act. |
| 2006 | 1381 | 15220 | 13815 | Fed raised rates. |
| 2007 | 1374 | 14602 | 14452 | Bank crisis. |
| 2008 | 1381 | 14535 | 14713 | Financial crisis. |
| 2009 | 1588 | 17160 | 14449 | Stimulus Act. |
| 2010 | 1680 | 17638 | 14992 | ACA. Dodd-Frank. |
| 2011 | 1859 | 18972 | 15543 | Japan earthquake. |
| 2012 | 2205 | 21778 | 16197 | Fiscal cliff. |
| 2013 | 2470 | 23723 | 16785 | Sequestration. |
| 2014 | 2688 | 24933 | 17522 | QE ends. |
| 2015 | 2938 | 26411 | 18219 | TPP. Iran deal. |
| 2016 | 3090 | 27263 | 18707 | Presidential race. |

*Source: Author´s calculations using data from* (Kimberly, 2019.; FinanciaForecastCenter 2019; OfficeforNationalStatistics, 2019; U.S.CensusBureau, 2019)

In Figure 1, there are visualized the time development of values of the Money Stock M1 in billions of $, M1 in billions of $ṁ, and GDP in billions of $ in the U.S.A. from 1960 through 2016.

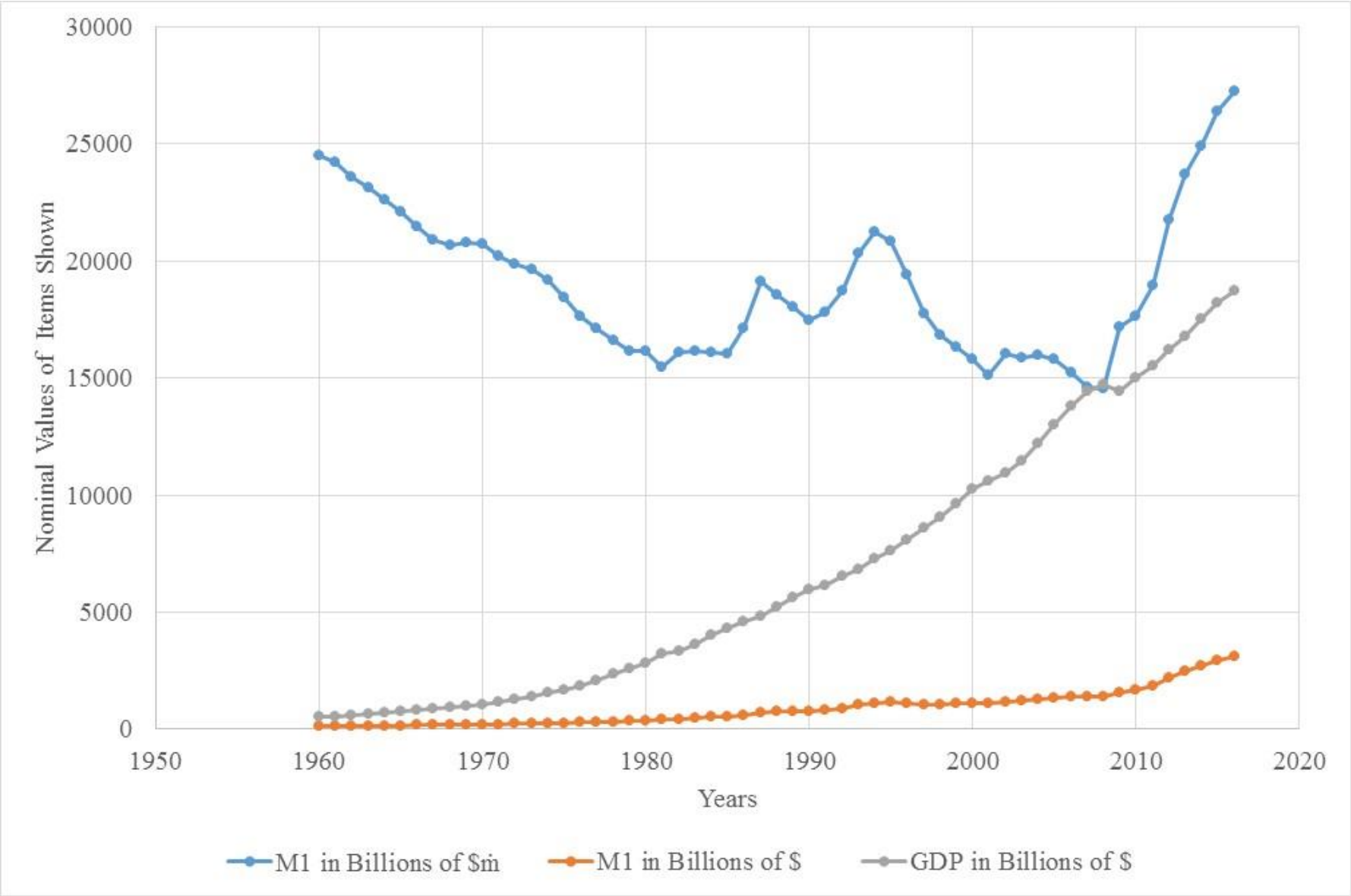

FIGURE 1. THE TIME DEVELOPMENT OF VALUES OF THE MONEY STOCK M1 IN BILLIONS OF $, M1 IN BILLIONS OF $Ṁ, AND GDP IN BILLIONS OF $ IN THE U.S.A. FROM 1960 THROUGH 2016.

*Notes:* The values used in Figure 1 are given in Table 5

*Source: Author´s calculations and design using data from* (FinanciaForecastCenter, 2019; Kimberly, 2019; OfficeforNationalStatistics, 2019; U.S.CensusBureau, 2019)

It is evident from the Table 5 and the Figure 1, that the amounts of money M1 and the values of GDP were mostly continuously growing from 1960 to 2016 – with exceptions when stagnated or was decreasing the M1 supply in 1995-2001, and 2006-2008 periods, and when appeared the GDP decrease in 2008-2009.

The course of the time dependence of M1 supply expressed in monetary minutes $ṁ showed a general decrease in the years 1960-2008, followed by a steep increase, which started in 2008). In the period of the general decrease, several peaks (in 1970, 1973, 1980, 1987, 1994, and 2002-2004) and through (in 1981, 1985, 90, 2001, and 2007-2008) cycles appeared.

## V. Discussion of the Results

This evaluation study is a part of my attempt to use Time as a standard (or commodity) for a new money system, which would be objective, predictable, dynamically evolving, and readily introduce-able in the economy, business, and everyday life.

In this work, I have postulated a time standard for a new Monetary Minute currency as the total Economic Time Capacity of a Year 525600 minutes. Consequently, I have set the Monetary Minute ṁ as a 1/525600 part of the total Economic Time Capacity of a Year. It has a Value of C*ṁ*, which is equal to a 1/525600 part of the GDP, p.c., expressed in a specific state currency C.

In the first part of the results, there is described how the Monetary Minutes *ṁ* were determined, and how their values C*ṁ* were calculated on the basis of the GDP and a number of population in specific economies, namely in the U.S.A., in the U.K., Germany, China, Japan and in the Czech Republic in this paper. The values found for *ṁ* expressed in the related state currencies*:* $*ṁ* 0.1210095 in the U.S.A., £*ṁ* 0.0146479 in the U.K., €*ṁ* 0.0777222 in Germany, ¥*ṁ* 0.0347098 in China, *Yṁ* 8.2872612 in Japan, *Kčṁ* 0.9519794 in the Czech Republic were mutually very different (see Table 1). It was obviously caused by different values of the GDP, numbers of population, consequently values of the GDP, p.c., different state currency value setting, and as well by different "aggregate productivity", i.e. different exploitation of the Total Time Capacity of a Year for generating of the GDP in Economies of individual states. The total 24 hours a day time of all the population, including the unemployed people, children, pensioners, rentiers, people with disabilities, etc. were counted into the Total Time Capacity of the Year. That means, the values of *ṁ* per se do not express the value of labor time itself. The ratio

between the labor and non-labor time capacity can be in principle very different in different countries with different political and social systems.

The values of the Monetary Minutes *ṁ*, expressed in the money markets were set by the actual valuation of related state currencies on the specific financial markets, too. In this work, the values of the C*ṁ* were: $*ṁ* 0.11918, £*ṁ* 0.0.170789, €*ṁ* 0.134971, ¥*ṁ* 0.00023033, *Yṁ* 0.1210095 according to the exchange rates of the Trading Economics (at July 1st, 2019, see Table 2).

The money market values of related C*ṁ* in Table 2 and the calculated values based on the GDP of the economies in Table 1 were significantly different. In my opinion, it illustrated the fact, that the values of currencies are set by the current situation (demand vs supply) in the financial markets rather than by the efficiency of the state economy.

The similar distortion of the values can be noticed in the commodity market in the Czech Republic. It was found, that prices of several commodities expressed in traditional currencies (USD, EUR, GBP) were very different and confusing, meanwhile, the ones expressed in the Monetary Minute currencies enabled to compare and testify their values concerning the currency market in the Czech Republic (see Table 3).

An overvaluation of the U.S. dollar against the values of other currencies in the Czech Republic at that time (at July 18th, 2019) could be derived from the MonMin ṁ values. Namely, the nominal values of one ounce of gold in the Czech commodity market, expressed in Monetary Minutes ṁ, were $ṁ 11958, Kčṁ 34603, €ṁ 16589, and £ṁ 79155, respectively. The equalization of the nominal Values of the prices of one ounce of gold in the Czech commodity market to a nominal Value of price of one ounce of gold $ṁ 11958 would be possible under a hypothetical presumption that the currency conversion ratios of the Czech Crown would be Kč/$ 7.867, Kč/€18.417, and Kč/£ 4.292, respectively (at July 18th, 2019).

In my concept, the Monetary Minutes *ṁ* does not serve as a gauge for measurement of Time amount, only, but they are used as a gauge (or a measure, or a scale) to measure, express, compare, etc. the quantity or "an amount" of Value of things, goods, services, processes, money, etc. via specific time units (*i.e. ṁ)*.

Thus, the nominal values of ṁ in Table 4 and Table 4b can be understood as a gauge (a measure) for measurement the relations between costs of products and a "consumption" of a human lifetime, as the values of costs were expressed in time units *ṁ* or in percentage of the salaries (which were also expressed in time units *ṁ* or in percent), which people received as a reward (a value) for their time they (ostensibly) spent, while generating their ideal share in the General Domestic Product. It can be noticed from the figures in the Table 4b, that the relative prices for meals and drinks in the countries under study were the same in order of several tenths to units of percent in 2019[5].

An interesting application of the Monetary Minute Currency ṁ could also be found in the regulation of M1 Stock. Namely, in Table 5 and Figure 1 was displayed, that the course of the time dependence of M1 supply expressed in monetary minutes $ṁ showed a general decreasing tendency in the years 1960-2008, followed by a steep increase, which started in 2008. In the period of the general decrease, several peaks (in 1970, 1973, 1980, 1987, 1994, and 2002-2004) and through (in 1981, 1985, 90, 2001, and 2007-2008) cycles appeared. The similar time dependence of the M1/GDP ratio was displayed in the classical monograph of

[5] The whole picture, namely the affordability of the food-staff products, goods, services, etc. should be, however, interpreted with caution – in relation with all of the family members, dependent on disposable salaries, unemployment rate, and the general social situation in specific countries.

Samuelson and Nordhaus (Samuelson and Nordhaus 2010, pp 176), which - in principle – have the same gist as the time dependence of monetary minutes ṁ. The abnormal dependences might reflect shortcomings in the generation of M1 supply in relation to GDP growth, and other interventions in tax and financial regulations, which were noted in Table 5, as well as the emergence of alternative sources of financing (like MBC, i.e. Math-Based Currency, like Bitcoin, Auroracoin, and hundreds of another ones) (Porter & Rousse, 2016). Especially, the displayed steep development of time dependence of Monetary Minutes ṁ, the number of which increased sharply over time, after 2008, its possible causes and consequences, deserve special utmost attention and analysis, in my opinion.

## VI. Conclusions

This study highlighted the reality, that traditional currencies, like USD, EUR, GBP, have different values depending on money regulator´s policies, financial market mood, exchange ratios policies, the access policies the international markets, specific products, commodities, and services market situations, purpose of their use, etc.

An attempt on how to improve the measurement of Values in Economy was designed in this work in the form of postulation a total Economic Time Capacity of a Year 525600 minutes as a time standard for a new Monetary Minute currency, definition the Monetary Minute ṁ as a 1/525600 part of the total Economic Time Capacity of a Year, and by determination of its Value (C*ṁ*) equal to a 1/525600 part of the GDP, p.c., expressed in a specific state currency C.

The study showed, that the Monetary Minute currency *ṁ* can serve for comparison of values, costs, and prices of the same or similar products, or - potentially - all economic entities expressed in different currencies, and to reveal disparities among them. It can be a prospective tool especially for tracing, measuring, and analyzing a wide spectrum of value transformations in the real economy, including production costs, efficiency, diversities in different countries or regions, and different historical eras in a more objective way than the current spectrum of currencies worldwide enable.

Monetary Minute currency ṁ cannot solve or change misstatements, indiscipline, greed, speculative behavior and violations of the basic principles and other shortages of generating and using money in the economy and society "on one´s own". Nevertheless, it might signalize, indicate, and quantify the speculative or unhealthy tendencies and relations among currency value settings, price settings, and in general Value settings.

The quantitative accuracy and reliability of the results given in this study were limited by the accuracy, reliability, and variability of some figures, namely on population, GDP values estimation, currency rates and another, available and the time of its creation.

Nevertheless, the quality of the fundamental finding, that the Monetary Minute currency ṁ can serve as a useful and practical tool for the Value determination, and for measurement of any economic entities (either element and/or processes) Values, seems to be undeniable.

I believe that the legitimacy and quality of the proposal for the new Monetary Minute currency will be subjected to the thorough broad verification from various points of view and possibly corrected.

Though I realize that the use of the hypothetical Monetary Minutes currency ṁ in a practical life will take much time, I offer a frame of it as a currency (Corporation, 2019) as follows:

*The Monetary Minute currency ṁ is based on time standard, which is defined as the total Economic Time Capacity of a Year 525600 minutes.*

*Its currency code is designed as MonMin, its currency symbol is designed as ṁ (read "min"), and its dimension is a Currency (in which the GDP, or the GDP, p.c. is expressed) per Monetary Minute.*

*The Central Bank for MonMin currency could be an authoritative recognized international institution. Future MonMin money might be issued in ṁ1, ṁ5, ṁ10, ṁ20, ṁ50, ṁ100, ṁ200, ṁ500, ṁ1000 denominations.*